\documentclass[%
 reprint,
 amsmath,assume,
 aps,
]{revtex4-2}
\usepackage{color}
\usepackage{here}
\usepackage{graphicx}
\usepackage{dcolumn}
\usepackage{bm}
\usepackage{float}
\usepackage{comment}
\usepackage[T1]{fontenc}

\begin{document}

\preprint{APS/123-QED}

\title{Single-Shot Magnetization Reversal in Ferromagnetic Spin Valves via Heat Control}

\author{Kazuaki Ishibashi$^{1,2,3}$}
\author{Junta Igarashi$^{3,4}$}
 \email{junta.igarashi@aist.go.jp}
\author{Alberto Anad\'{o}n$^{3}$}
\author{Michel Hehn$^{3,5}$}
\author{Yann Le Guen$^{3}$}
\author{Satoshi Iihama$^{6}$}
\author{Julius Hohlfeld$^{3}$}
\author{Jon Gorchon$^{3}$}
\author{St\'{e}phane Mangin$^{3,5}$}
\author{Gr\'{e}gory Malinowski$^{3}$}

\affiliation{$^1$Department of Applied Physics, Graduate School of Engineering, Tohoku University, Sendai 980-8579, Japan}
\affiliation{$^2$WPI Advanced Institute for Materials Research (AIMR), Tohoku University, 2-1-1, Katahira, Sendai 980-8577, Japan}
\affiliation{$^3$Universit\'{e} de Lorraine, CNRS, IJL, Nancy, France}
\affiliation{$^4$National Institute of Advanced Industrial Science and Technology, Tsukuba 305-8563, Japan}
\affiliation{$^5$Center for Science and Innovation in Spintronics (CSIS), Core Research Cluster (CRC), Tohoku University, Sendai 980-8577, Japan}
\affiliation{$^6$Department of Materials Physics, Nagoya University, Nagoya 464-8603, Japan}

\date{\today}

\begin{abstract}
We study laser-induced ultrafast magnetization reversal in a ferromagnetic spin valve by comparing the effects of direct laser excitation and ultrashort hot-electron pulses. A wedged Cu layer is deposited atop the spin valve to tune energy transmission to the magnetic stack for both optical and hot-electron excitation. We demonstrate single-shot magnetization reversal of the free layer using hot-electron pulses. Moreover, such reversal is achieved even with picosecond laser pulses. The influence of laser fluence, Cu thickness ($t_{\mathrm{Cu}}$), and pulse duration is investigated in detail. Our results indicate that the key factor enabling magnetization reversal is full demagnetization of the free layer, driven by a rapid rise in its electronic temperature—achieved via either direct laser or hot-electron excitation. This work advances the understanding of ultrafast magnetization reversal via nonlocal heat and spin transport under strongly out-of-equilibrium conditions.

\end{abstract}
 
\maketitle


\textit{Introduction.} Ultrafast stimuli, such as a femtosecond laser pulse, have been shown to bring magnetic materials into an out-of-equilibrium state, resulting in ultrafast demagnetization \cite{beaurepaire1996ultrafast}. Moreover, certain ferrimagnet systems, such as gadolinium-transition metal (Gd-TM)-based alloys and multilayers, as well as MnRuGa, undergo an ultrafast magnetization reversal upon single femtosecond laser pulse excitation \cite{radu2011transient,ostler2012ultrafast,lalieu2017deterministic,banerjee2020single,davies2020exchange,banerjee2021ultrafast}. In Gd-TM systems, ultrafast magnetization reversal is mediated by angular momentum transfer between the two antiferromagnetically exchange-coupled Gd and TM sublattices. 

In 2018, Iihama \textit{et al.} demonstrated the magnetization reversal of a ferromagnetic layer (FM) in a GdFeCo/Cu/FM spin-valve structure \cite{iihama2018single}. An angular momentum transfer from the ferrimagnetic alloy (GdFeCo) to the FM layer, mediated by a spin current through the Cu spacer layer, was shown to be responsible for the FM layer reversal. From these measurements, the reversal process was shown to be compatible with a spin current originating from the demagnetization of Gd \cite{iihama2018single,remy2020energy}.

In 2023, Igarashi \textit{et al.} demonstrated subpicosecond magnetization reversal in Gd-free [Co/Pt]/Cu/[Co/Pt] spin valves using a single laser pulse, structures typically used for spin-transfer torque switching \cite{slonczewski1996current, berger1996emission, mangin2006current, igarashi2023optically}. The switching mechanism differs from that of Gd-based systems. Indeed, as described in Refs. \cite{igarashi2023optically,igarashi2023inter}, the injection of opposite-sign spin current into the free layer must be taken into account in order to explain the spin valve magnetization switching from parallel to antiparallel alignment. Although the microscopic mechanism remains unclear, two scenarios based on spin-current-mediated switching have been proposed: (i) spin current generated by demagnetization of the free layer, and (ii) that from the reference layer \cite{igarashi2023optically}. In case (i), ultrafast demagnetization in the free layer generates a spin current that travels through the Cu spacer, reflects and flips at the reference interface, and returns to switch the free layer. While this reflection model qualitatively reproduces experiments \cite{remy2023ultrafast}, further refinement is needed.

One fundamental question, both for applications and for understanding the physical mechanism behind these results, is whether light is necessary or if other heat-inducing stimuli can produce similar results. Indeed, an ultrashort hot-electron pulse, generated by shining light on a non-magnetic metallic layer, have been shown to efficiently carry heat and induce ultrafast demagnetization of an adjacent FM layer \cite{bergeard2016hot,bergeard2020tailoring,pudell2020heat}, as well as induce full magnetization reversal in ferrimagnetic GdFeCo \cite{wilson2017ultrafast,xu2017ultrafast}.

In this Letter, we investigate magnetization reversal using both a femtosecond light pulse and a hot-electron pulse in [Co/Pt]$_{3}$/Co/Cu(10)/[Co/Pt]$_{2}$/Cu($t_{\mathrm{Cu}}$)/capping heterostructures. The effects of light and hot electrons can be tuned by varying the Cu thickness ($t_{\mathrm{Cu}}$), the nature of the capping, and the side of the sample illuminated by the laser pulse.

\begin{figure}
    \centering
    \includegraphics[width=1\linewidth, clip]
    {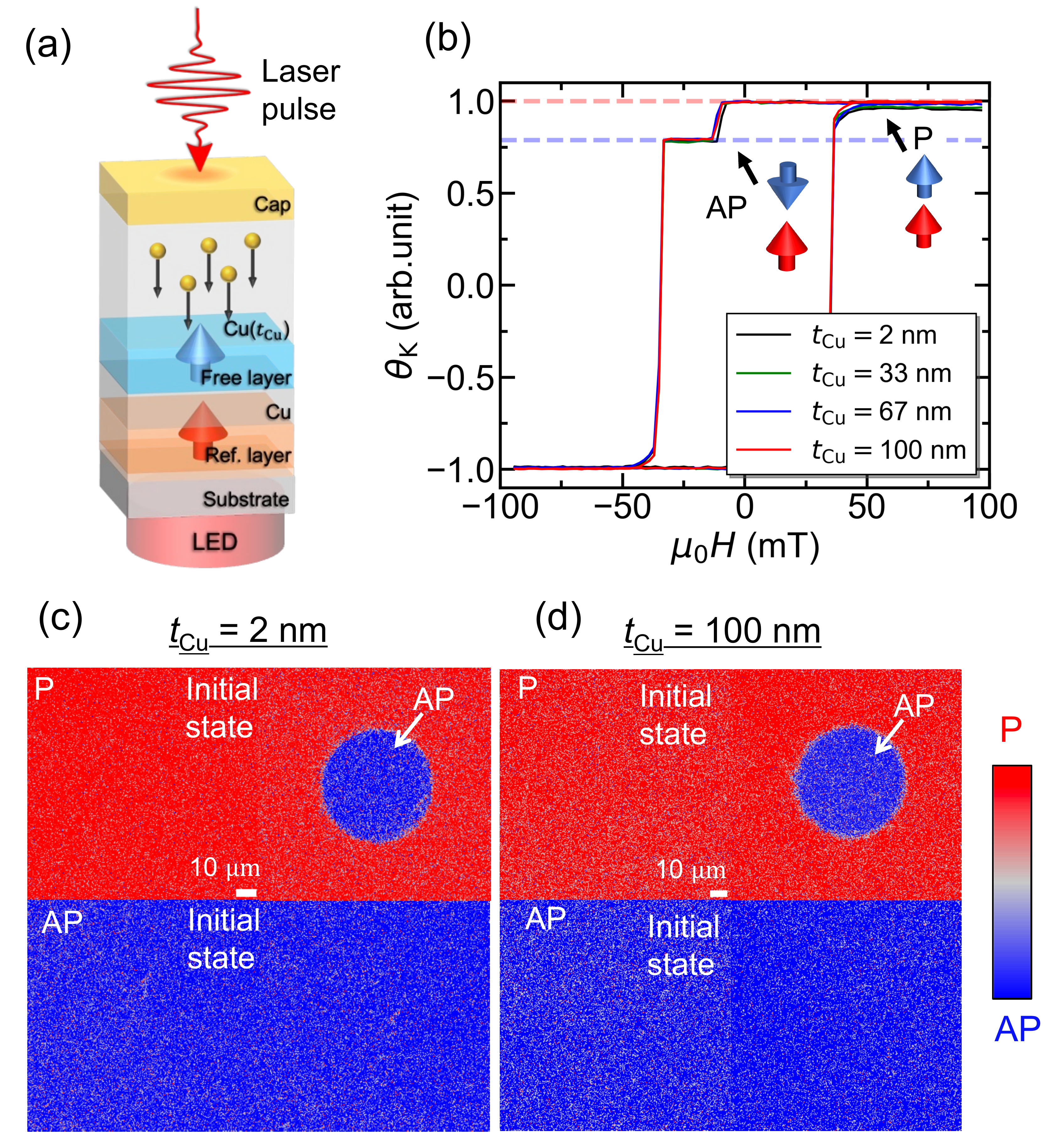}
    \caption{(a) Sample structure and geometry. MOKE images were taken using a 628 nm LED. (b) MOKE hysteresis loops for different $t_\mathrm{Cu}$. (c, d) MOKE images before and after laser irradiation for $t_\mathrm{Cu} = 2$ nm ($F = 9.3$ mJ/cm$^2$) and 100 nm ($F = 64$ mJ/cm$^2$), respectively.}
    \label{fig1}
\end{figure}

\textit{Sample stack and experimental methods.} All samples were grown by sputtering. The basic stack structure is Ta(5)/Pt(4)/[Co(0.82)/Pt(1)]$_{3}$/Co(0.82)/\allowbreak Cu(10)/[Co(0.49)/Pt(1)]$_{2}$/Cu($t_{\mathrm{Cu}}$)/Ta(5) deposited on a glass substrate (thicknesses are in nm). All samples have the same spin valve structure with a 10-nm-thick Cu spacer between the two [Co/Pt] multilayers. We define the bottom (i.e., closer to the substrate) and top (i.e., farther from the substrate) [Co/Pt] multilayers as the reference and free layers, respectively. The reference layer has a higher Curie temperature than the free layer. A Cu layer was deposited on top of the spin valve to tune the optical absorption, as demonstrated in previous studies \cite{bergeard2016hot,bergeard2020tailoring}. The thickness of this Cu layer, $t_{\mathrm{Cu}}$, was varied from 2 to 100 nm using a wedge deposition method \cite{peng2023plane}, corresponding to three sample groups: 2 nm, 10–30 nm, and 33–100 nm. Figure \ref{fig1}(a) shows a schematic illustration of the sample stack and typical experimental geometry. A linearly-polarized laser pulse was used for our experiments. The laser pulse was generated from a Ti:Sapphire femtosecond laser source with a wavelength of 800 nm and a repetition frequency of 5 kHz. The pulse duration was varied from 50 fs to 10 ps. A light-emitting diode (LED) source with a center wavelength of 628 nm was used to capture magneto-optical Kerr effect (MOKE) images of the samples \cite{iihama2018single}. We pumped from the capping layer or the substrate side and always probed from the substrate side to measure the magnetic response, for all Cu thicknesses.  Figure \ref{fig1}(b) shows hysteresis loops obtained via static MOKE measurements with an external magnetic field applied perpendicular to the film. We confirmed a perpendicular easy axis in both layers and four magnetic configurations. In this study, we focus on the P and AP states indicated in Fig. \ref{fig1}(b). The coercive field of both the free and the reference layers remained the same regardless of $t_{\mathrm{Cu}}$.

\textit{Single-shot experiment by shining a laser pulse on the capping layer.} First, we investigate single-shot switching by shining a laser pulse on the capping layer, as illustrated in Fig. \ref{fig1}(a). Figure \ref{fig1}(c) shows MOKE images taken before and after laser excitation starting from P (red) and AP (blue) states for $t_{\mathrm{Cu}}$ = 2 nm. For such a thin Cu layer, the laser pulse directly reaches the spin valve, leading to direct laser-induced demagnetization. In this case, a clear magnetization reversal from P to AP state (P-to-AP switching) is observed as reported in previous works \cite{igarashi2023optically,igarashi2023inter}. Figure \ref{fig1}(d) shows the results of the single-shot experiment for $t_{\mathrm{Cu}}$ = 100 nm, wherein the laser pulse cannot reach the spin valve. Nevertheless, P-to-AP switching is still observed, indicating that hot electrons generated from the capping layer and flowing into the free layer are sufficient to trigger P-to-AP switching in the ferromagnetic spin valves. In contrast, AP-to-P switching was not observed in the studied samples. As previous studies have shown that reducing the thickness of the Cu spacer improves AP-to-P switching \cite{igarashi2023optically,igarashi2023inter}, we henceforth focus solely on P-to-AP switching. 

\begin{figure}
    \centering
    \includegraphics[width=0.7\linewidth, clip]
    {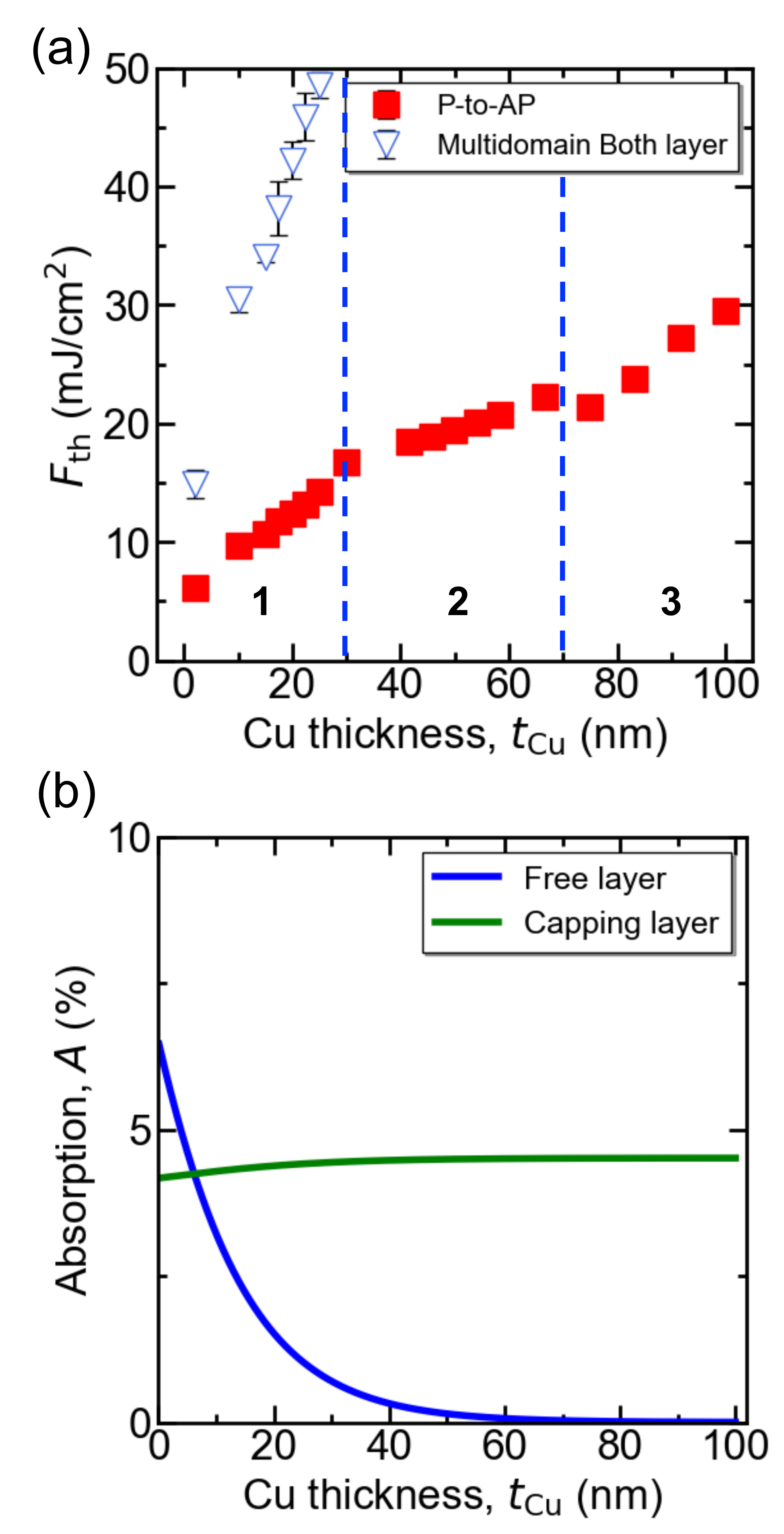}
    \caption{Summary of the single-shot switching experiments in Ta(5)/Pt(4)/[Co(0.82)/Pt(1)]$_{3}$/Co(0.82)/Cu(10)\allowbreak /[Co(0.49)/Pt(1)]$_{2}$/Cu($t_{\mathrm{Cu}}$)/Ta(5). (a) Evolution of the threshold fluence for P-to-AP switching ($F_{\text{P}}$) and multidomain state ($F_{\text{MD}}$) as a function of $t_{\mathrm{Cu}}$. (b) Calculation of laser energy absorption in the free and capping layers as a function of $t_{\mathrm{Cu}}$.}
    \label{fig2}
\end{figure}

Figure \ref{fig2}(a) summarizes threshold fluences $F_{\mathrm{th}}$ for P-to-AP switching ($F_{\mathrm{P}}$) and for multidomain state ($F_{\mathrm{MD}}$). The value of $F_{\mathrm{th}}$ is determined for each state by fitting the domain area as a function of various laser pulse energies, as described in the Supplementary Material of Ref. \cite{igarashi2023inter}. Both $F_{\mathrm{P}}$ and $F_{\mathrm{MD}}$ are observed to increase with $t_{\mathrm{Cu}}$. The change in slope of $F_{\mathrm{P}}$ defines three different regimes (1, 2, and 3), as indicated in Fig. \ref{fig2}(a). The existence of these regimes can be associated to differences in energy absorption, which is calculated using a transfer matrix method and refractive indices for each material collected from Refs. \cite{igarashi2020engineering,johnson1974optical}. Optical energy absorption in the free layer, shown in Fig. \ref{fig2}(b), decreases significantly with increasing $t_{\mathrm{Cu}}$ and reaches less than 0.1\% at $t_{\mathrm{Cu}}$ = 60 nm. From both experiments and calculations, we identify three distinct regions. In region 1 ($t_{\mathrm{Cu}} < 30 \, \mathrm{nm}$), both direct optical excitation and hot-electron-induced excitation contribute to magnetization switching. In region 1, the optical absorption in the free layer remains significant but decreases rapidly with increasing Cu thickness, resulting in a steep slope. In region 2 (30--70 nm), hot-electron excitation dominates, although some optical contribution may remain. The relatively shallow slope in this region is attributed to the long mean free path of hot electrons in Cu, which enables efficient transmission through the Cu layer, as reported previously \cite{bauer2015hot, bergeard2016hot}. In addition, the smaller change of the absorption in the free layer also leads to the relatively gentle slope. In region 3 ($t_{\mathrm{Cu}} > 70 \,\mathrm{nm}$), indirect excitation contributes dominantly to the switching. The slope increases again and becomes comparable to that in region 1. This behavior is somewhat unexpected, as it deviates from our absorption calculations (Fig. \ref{fig2}(b)) and the anticipated long mean free path of hot electrons in Cu. One possible explanation is that increasing $t_{\mathrm{Cu}}$ leads to increased interfacial roughness between the capping layer and the Cu layer, which in turn enhances electron scattering at the interface. As a result, fewer hot electrons reach the free layer, requiring higher laser fluence to achieve switching.

We also investigate how the capping layer affects P-to-AP switching by testing different capping materials: Ta, Pt, and no capping. The lowest $F_{\mathrm{P}}$ is observed at approximately $t_{\mathrm{cap}}$ = 7 to 10 nm, which represents the optimal compromise between laser absorption depth and hot-electron scattering within the capping layer itself \cite{bergeard2020tailoring}. For further details on capping layer dependence, see End Matter.

\begin{figure}[t]
    \centering
    \includegraphics[width=1\linewidth, clip]
    {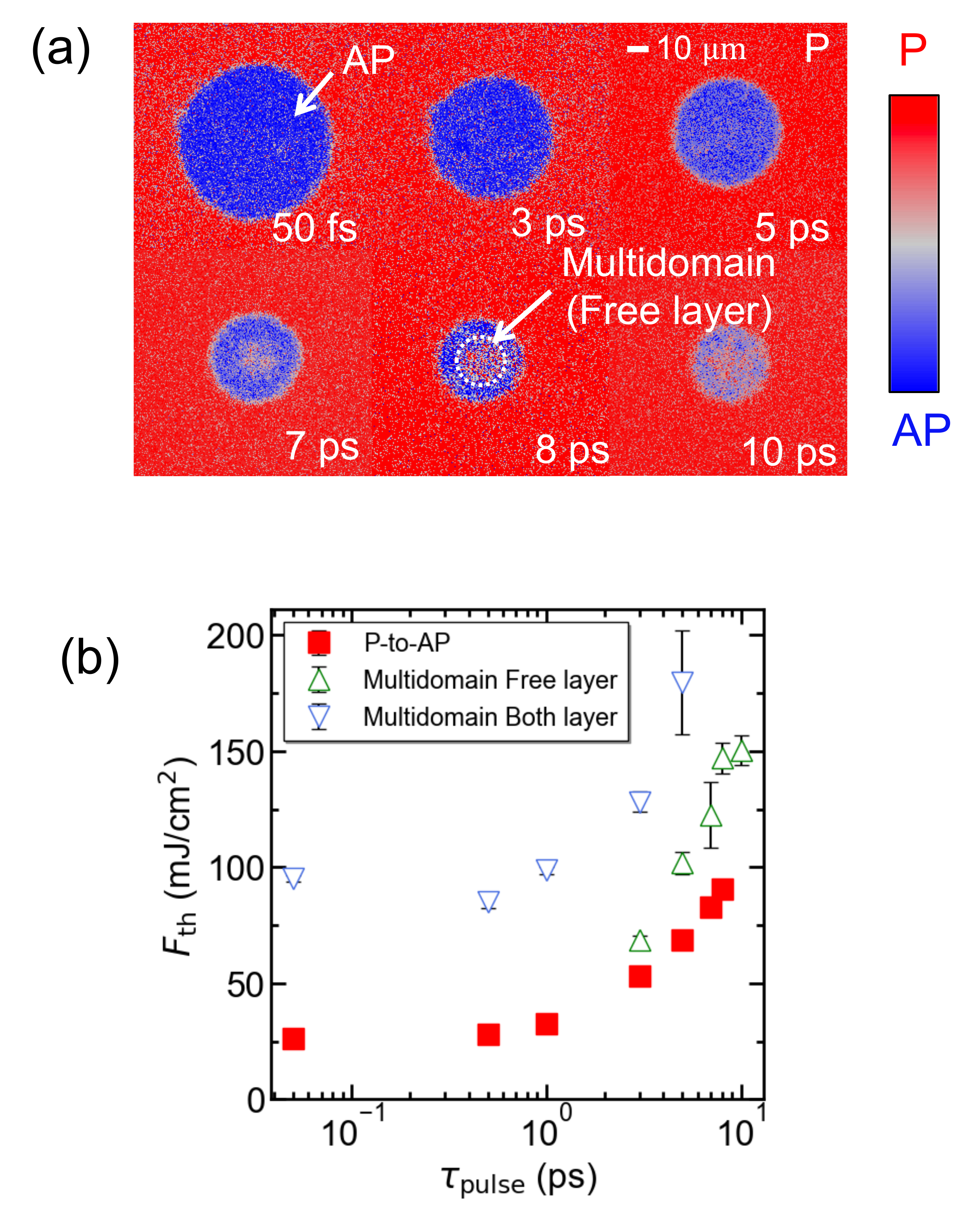}
    \caption{(a) MOKE images after laser pulses with various durations in the structure with $t_\mathrm{Cu}=100$ nm. (b) Threshold fluence $F_\mathrm{p}$ and $F_\mathrm{MD}$ vs pulse duration.}
    \label{fig3}
\end{figure}

\begin{figure*}[t]
    \centering
    \includegraphics[width=1\linewidth, clip]{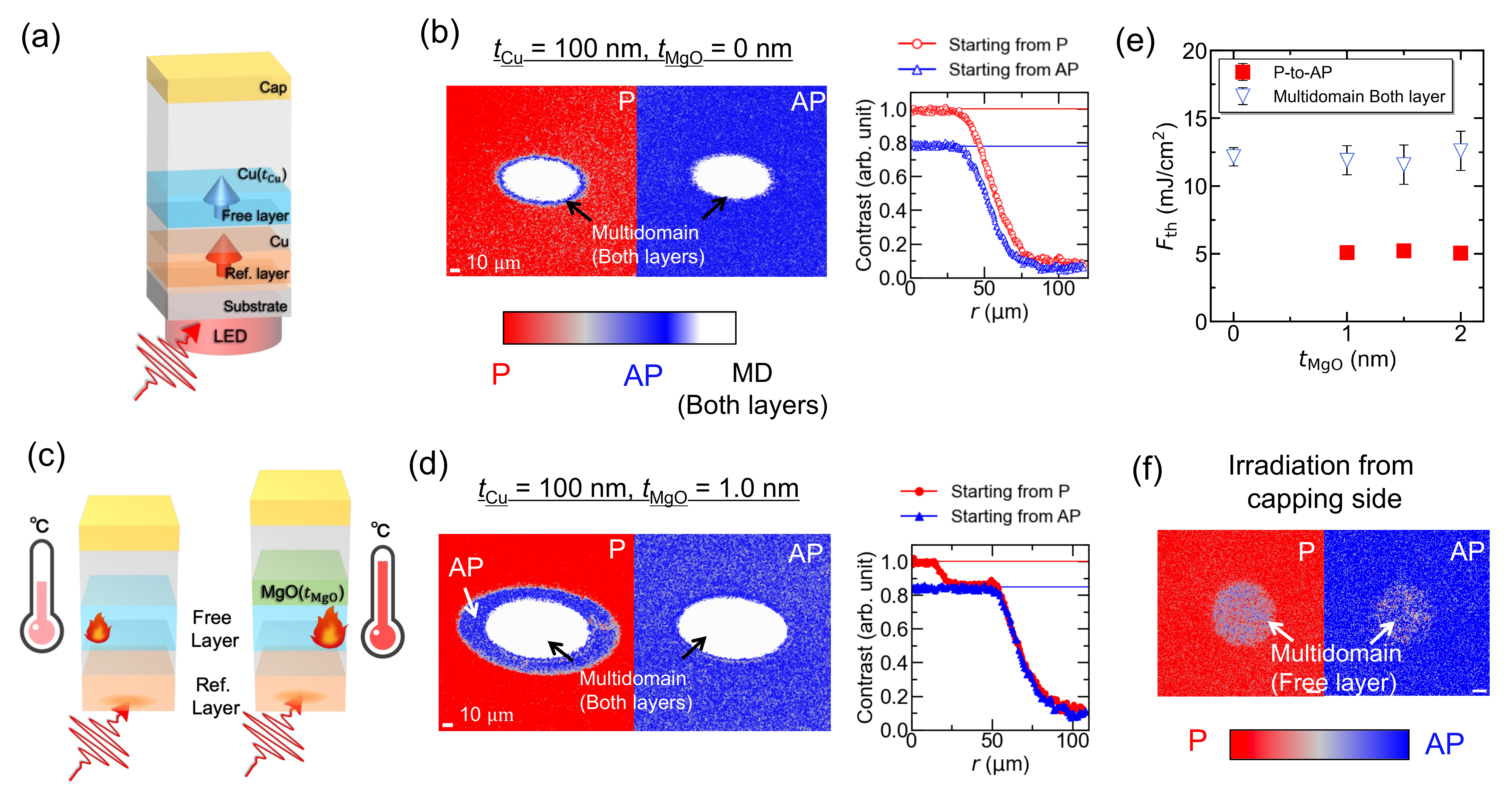}
    \caption{(a) Schematic illustration of the experimental geometry. (b,d) MOKE images and line profiles after a laser pulse for $t_\mathrm{MgO}=0$ and $1$ nm. (c) Schematic illustration of sample structure and the laser geometry. (e) $F_\mathrm{p}$ and $F_\mathrm{MD}$ as a function of $t_\mathrm{MgO}$. (f) Image for $t_\mathrm{MgO} = 1$ nm irradiated from the capping layer ($F = 137$ mJ/cm$^2$).}
    \label{fig4}
\end{figure*}
In the following, we investigate the effect of pulse duration $\tau_{\mathrm{pulse}}$ on P-to-AP switching threshold fluence, which can provide valuable insight into all-optical switching (AOS)  \cite{wei2021all,peng2023plane}. In a previous study, on a similar [Co/Pt]/Cu(10)/[Co/Pt] spin valve, P-to-AP switching was observed up to $\tau_{\mathrm{pulse}}$ = 1 ps, while a multidomain state in the free layer was observed instead for longer pulses \cite{igarashi2023optically}. The Cu layer on top of GdFeCo has been demonstrated to act as a heat sink, preventing long-term heat accumulation in GdFeCo and thus extending the pulse duration over which AOS can be observed \cite{verges2024extending}. From this, we expect P-to-AP switching to occur for extended pulse durations in our samples as well. Figure \ref{fig3}(a) shows MOKE images for various pulse durations taken after shining a laser pulse from the capping layer in Ta(5)/Pt(4)/[Co(0.82)/Pt(1)]$_{3}$/Co(0.82)/\allowbreak Cu(10)/[Co(0.49)/Pt(1)]$_{2}$/Cu(100)/Ta(5). P-to-AP switching can be observed with pulses up to $\tau_{\mathrm{pulse}}$ = 8 ps, due to the Cu heat sink. Furthermore, a multidomain state in the free layer appears in the center of the beam for longer pulse durations, such as 7, 8, and 10 ps. This confirms that the Cu heat sink extends the switching window, in line with prior GdFeCo/Cu results \cite{verges2024extending}. Note that this provides the potential of achieving the switching induced by heat pulse using picosecond laser or even electrical pulse \cite{yang2017ultrafast}.  Figure \ref{fig3}(b) summarizes $F_{\mathrm{P}}$ and $F_{\mathrm{MD}}$, the latter for both free and reference layer. Both $F_{\mathrm{P}}$ and $F_{\mathrm{MD}}$ increase with pulse duration, as previously reported in Ref. \cite{igarashi2023optically} in samples that did not include a heat sink. As the pulse duration increases, the peak in electronic temperature decreases, while the phonon temperature remains unchanged \cite{peng2023plane}. A higher laser fluence is required to induce sufficient demagnetization within a short timescale. Consequently, the threshold fluence increases as the pulse duration becomes longer. These results suggest that the rise in electronic temperature, which enables substantial demagnetization of the free layer, plays a crucial role in the P-to-AP switching of ferromagnetic spin valves.

\textit{Single-shot experiment shining on the reference layer.} In order to compare the effect of direct and indirect excitation on P-to-AP switching in the same sample, the sample is irradiated with a laser pulse from the reference layer as illustrated in Fig. \ref{fig4}(a). Figure \ref{fig4}(b) shows MOKE images obtained after laser excitation for samples with $t_{\mathrm{Cu}}$ = 100 nm. Surprisingly, no P-to-AP switching is observed despite having observed P-to-AP switching in the same sample with a laser excitation from the capping side. It should be noted that the blue ring in Fig. \ref{fig4}(b) (left) does not correspond to switching, as indicated by the line profile shown in the right side. As shown in End matter, P-to-AP switching could be observed in $t_{\mathrm{Cu}}$ up to 20 nm.  

To further investigate the switching mechanism, we inserted an insulating MgO layer between the free layer and 100-nm-thick Cu, as shown in Fig.~\ref{fig4}(c). MgO suppresses electron transport, blocking both spin and heat currents \cite{wahada2022atomic, rouzegar2023terahertz, jang2020thermal}. Figure~\ref{fig4}(d) shows MOKE images after laser irradiation for the sample with MgO. Conversely, the same sample with an inserted MgO layer shows P-to-AP switching as indicated by the line profile shown in right side of Fig. \ref{fig4}(d). Figure \ref{fig4}(e) summarizes $F_{\mathrm{th}}$ as a function of MgO thickness $t_{\mathrm{MgO}}$, where it can be observed that $F_{\mathrm{th}}$ does not depend on $t_{\mathrm{MgO}}$. Note that we observe only P-to-AP switching with the laser fluence between $F_{\mathrm{P}}$ and $F_{\mathrm{MD}}$ for the sample with MgO layer.

To understand this behavior, we perform two-temperature model calculations \cite{hohlfeld2000electron, pudell2020heat}. The results show that the electronic temperature of the free layer is higher with MgO than without, while that of the reference layer remains unchanged. This suggests electron accumulation in the free layer due to the MgO barrier, leading to heat retention and increased temperature, as illustrated in Fig.~\ref{fig4}(c). This efficient rise in electronic temperature may facilitate the ultrafast demagnetization of the free layer, which in turn enhances the P-to-AP switching process. The details of the calculation are described in End Matter. 

We also investigate the effect of MgO when irradiating from the capping side. Figure~\ref{fig4}(f) shows MOKE images for $t_{\mathrm{MgO}} = 1$ nm. No P-to-AP switching is seen, even at high fluence ($F > 100$ mJ/cm$^2$), though a multidomain state of the free layer appears. This suggests that 1-nm MgO strongly blocks electron transport. The multidomain may arise from heat transport via phonons in MgO, resulting in long-term heating of the free layer \cite{pudell2020heat, mattern2023concepts}.

\textit{Discussion.} We demonstrate that controlling the electronic temperature in the free layer plays a crucial role in enabling P-to-AP switching. Our results show that switching can be triggered not only by direct optical excitation but also by indirect excitation via hot electrons.

Regarding the mechanism, we consider that P-to-AP switching is driven by spin-current injection into the free layer, consistent with previous reports \cite{igarashi2023optically, remy2023ultrafast}. However, the microscopic origin of this spin current remains unresolved. As noted in the Introduction, two scenarios are possible: (i) spin current reflected from the interface following ultrafast demagnetization of the free layer, and (ii) spin current emitted from the reference layer during its remagnetization.

In this study, we focused on identifying the conditions needed for switching. We find that full demagnetization of the free layer—whether induced by direct or indirect excitation—is sufficient to trigger P-to-AP reversal. Notably, our results show that ultrafast demagnetization, essential for generating spin currents in both scenarios\cite{igarashi2023optically, remy2023ultrafast}, can be induced by thermal stimuli such as a hot-electron pulse. This indicates that light is not essential; heat alone suffices to drive the required spin dynamics. Also, since both scenarios require the presence of both the free and reference layers, we do not expect to observe P-to-AP switching in samples lacking the reference layer, as reported in a previous study using direct optical excitation \cite{igarashi2023optically}. 

Time-resolved measurements of hot-electron-induced magnetization dynamics could not be performed due to experimental challenges, such as isolating the magneto-optical signal from the free layer. However, prior studies on [Co/Pt] multilayers have shown that hot-electron-induced demagnetization occurs on ultrafast timescales similar to those of direct optical excitation \cite{bergeard2016hot, bergeard2020tailoring}. Based on these findings, we expect that hot-electron-induced P-to-AP switching in our system occurs on comparable ultrafast timescales, as reported in our earlier work \cite{igarashi2023optically}. 

Distinguishing between the two spin-current generation mechanisms remains experimentally challenging and is beyond the scope of this work.

While our findings clarify the overall switching mechanism, one point remains open: P-to-AP switching occurs even without an MgO layer when the laser is incident from the capping-layer side. Given the thermal gradient in this configuration, one possible explanation is that the free layer is more efficiently heated from the top, leading to a higher electron temperature than with excitation from the substrate side. Further investigation is needed to fully clarify this effect.

\textit{Conclusion.} We demonstrate single-shot switching in Co/Pt spin valves without requiring direct laser excitation. Introducing a Cu heat sink layer extends the pulse duration window for observing P-to-AP switching.

Our experiments reveal that controlling the electronic temperature—via either direct optical or indirect hot-electron excitation—is crucial to the switching mechanism. These findings provide new insight into ultrafast magnetization reversal driven by nonlocal spin transport under strongly nonequilibrium conditions, and suggest that switching can be realized using more practical sources, such as picosecond laser or even electrical pulses \cite{yang2017ultrafast}.

\textit{Acknowledgements.} We thank E. D\'{i}az for fruitful discussion. This work is supported by the French National Research Agency (ANR) through the France 2030 government PEPR Electronic grants EMCOM (ANR-22-PEEL-0009) by the ANR SPOTZ project ANR- 20-CE24- 0003, SLAM ANR- 23-CE30-0047, the project MAT-PULSE from “Lorraine Université d’Excellence” reference ANR-15- IDEX-04-LUE, the Institute Carnot ICEEl, the Région Grand Est, the Metropole Grand Nancy, the “FEDERFSE Lorraine et Massif Vosges 2014-2020”, a European Union Program,  the Academy of Finland (Grant No. 316857), the Sakura Program, the JSPS Bilateral Program, JSPS KAKENHI JP 24K22964, the Tohoku University-Universite de Lorraine Matching Funds, and CSIS cooperative research project in Tohoku University. This article is based upon work from COST Action CA23136 CHIROMAG, supported by COST (European Cooperation in Science and Technology). K. I. acknowledges Grant-in-Aid for JSPS Fellow (No. 22J22178), JST ASPIRE (No. JPMJAP2409), X-NICS, and GP-Spin at Tohoku University. J.I. acknowledges support from JSPS Overseas Research Fellowships. S. I. acknowledges Grant-in-Aid for Transformative Research Areas (No. 24H02235). All funding was shared equally among all authors. 

\nocite{*}

\bibliography{Hotelectron}

\section*{End Matter}

\begin{figure}[t!]
    \centering
    \includegraphics[width=0.8\linewidth, clip]
    {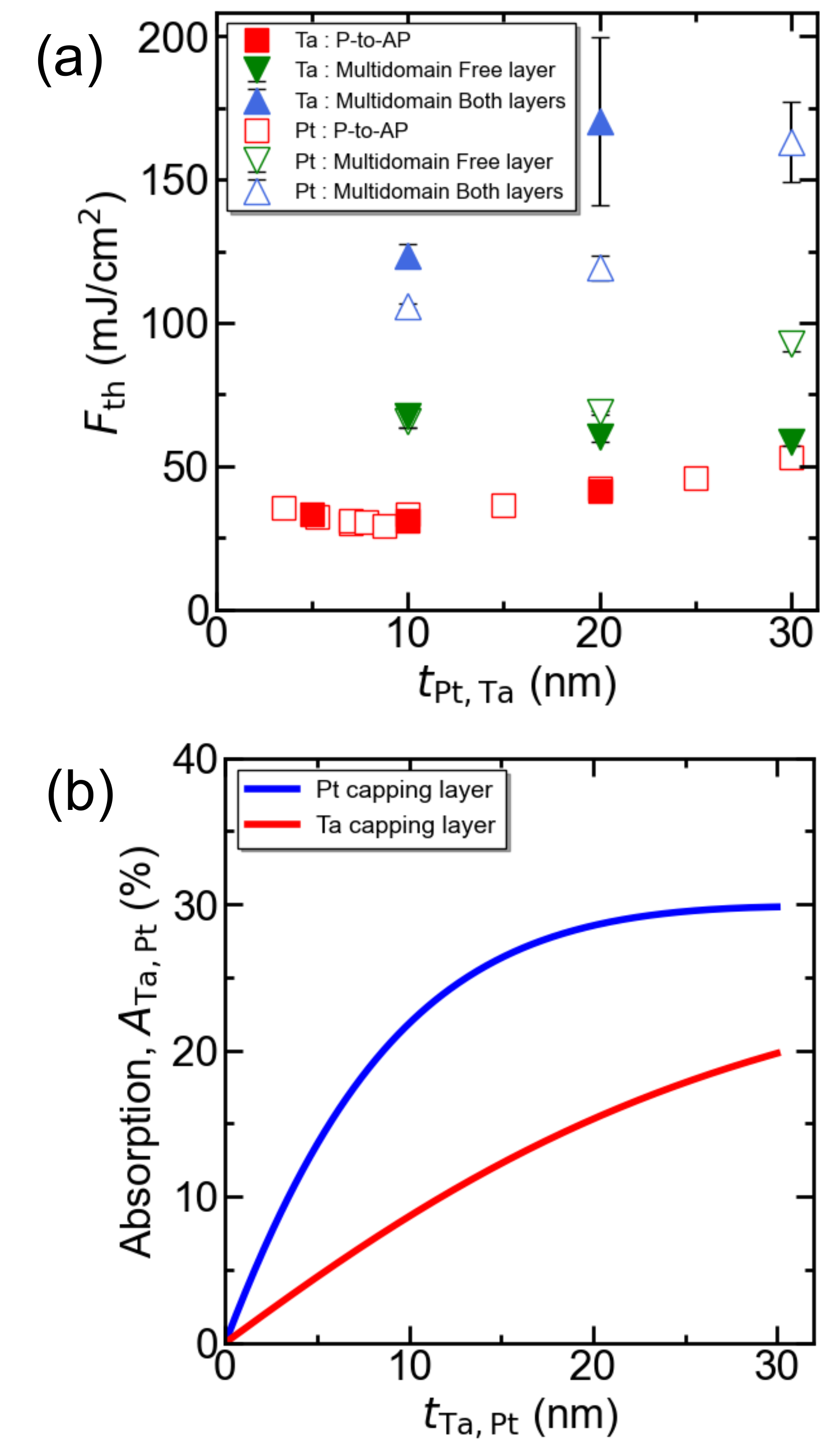}
    \caption{Single-shot experiment in Ta(5)/Pt(4)/[Co(0.82)/Pt(1)]$_{3}$/Co(0.82)/Cu(10)\allowbreak/[Co(0.49)/Pt(1)]$_{2}$ /Cu(100)/Cap($t_{\mathrm{cap}}$). The sample is irradiated with a laser pulse from the capping layer as shown in Fig. \ref{fig1}(a). (a) Evolution of the threshold fluences as a function of capping layer thickness. (b) Calculated laser energy absorption in the capping layer as a function of the capping layer thickness.}
    \label{EM_fig1}
    \centering
\end{figure}

\textit{Appendix A: Influence of the capping layer on switching.} To investigate how the capping layer affects P-to-AP switching in ferromagnetic spin valves, we prepare samples with different cappings: Ta, Pt, and no capping. $t_{\mathrm{Cu}}$ was fixed to 100 nm. The capping layer thickness $t_{\mathrm{cap}}$ was varied from 5 to 30 nm for Ta. For Pt, $t_{\mathrm{cap}}$ was varied linearly from 3.5 to 10.5 nm and from 10 to 30 nm using a wedge deposition technique \cite{peng2023plane}. Figure \ref{EM_fig1}(a) summarizes $F_{\mathrm{th}}$ for various cappings. No switching could be observed in the absence of a capping, which indicates that naked Cu generates hot electrons less efficiently compared with Ta- and Pt-capped Cu. $F_{\mathrm{P}}$ shows a similar trend regardless of materials: the lowest $F_{\mathrm{P}}$ is observed at approximately for $t_{\mathrm{cap}}$ between 7 and 10 nm, which has been demonstrated to be the best compromise between laser absorption depth and hot electron scattering in the capping layer itself \cite{bergeard2020tailoring}. Figure \ref{EM_fig1}(b) shows the energy absorption as a function of $t_{\mathrm{cap}}$ for the Pt and Ta cappings. In the calculation, Pt absorbs the energy of the laser pulse to a greater extent than Ta, with a factor of more than 1.5, but nevertheless, the value of $F_{\mathrm{P}}$ is almost the same between Pt and Ta. Conversely, $F_{\mathrm{MD}}$ shows different behavior for Pt and Ta: it increases with $t_{\mathrm{Pt}}$ while it decreases with $t_{\mathrm{Ta}}$. 

It should be noted that while the calculation considers pure Ta, the actual Ta capping is oxidized, which may contribute to discrepancies between the experimental and calculated results.

\begin{figure}
    \centering
    \includegraphics[width=0.8\linewidth, clip]
    {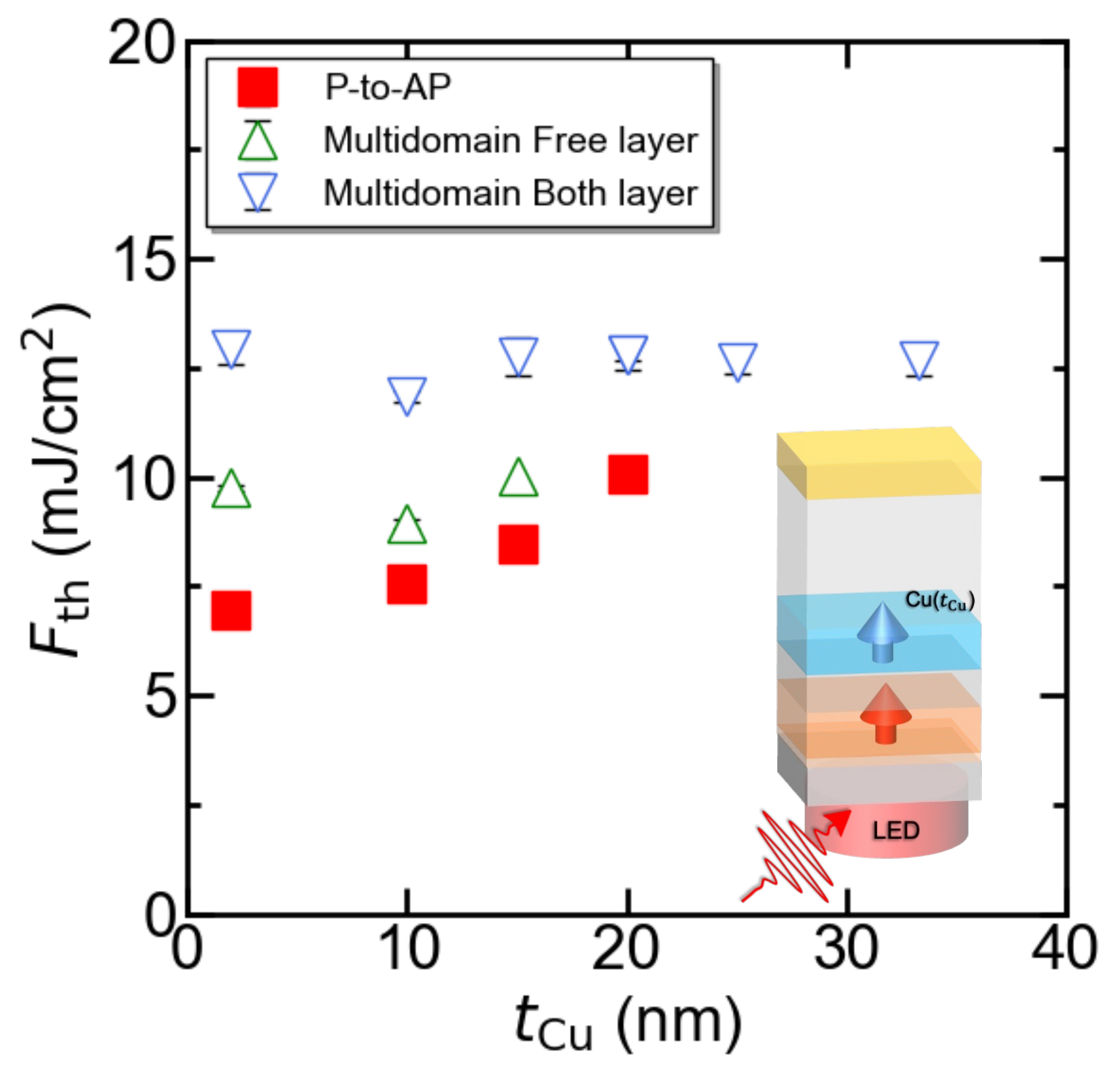}
    \caption{Single-shot experiment in Ta(5)/Pt(4)/[Co(0.82)/Pt(1)]$_{3}$/Co(0.82)/Cu(10)/\allowbreak [Co(0.49)/Pt(1)]$_{2}$/Cu($t_{\mathrm{Cu}}$)/Ta(5). The sample is irradiated with a laser pulse from the reference layer as shown in Fig. \ref{fig4}(a). Evolution of the threshold fluences as a function of Cu thickness. }
    \label{EM_fig2}
    \centering
\end{figure}

\textit{Appendix B: Cu thickness dependence of switching with irradaiation from substrate side.}Although P-to-AP switching was clearly observed in the same sample with the laser irradiation from capping side, 100-nm-thick Cu sample exhibits no P-to-AP switching in case of excitation from substrate side as mentioned above. To investigate the influence of the irradiation from substrate side on the switching in detail, Cu thickness was varied as shown in Fig. \ref{fig4}(a). Figure \ref{EM_fig2} shows evolution of threshold fluences plotted as a function of Cu thickness. P-to-AP switching is observed for $t_{\mathrm{Cu}}$ = 2 nm. Conversely, no P-to-AP switching is observed for $t_{\mathrm{Cu}}$ = 25 nm. $F_{\mathrm{P}}$ increases with $t_{\mathrm{Cu}}$ while $F_{\mathrm{MD}}$, for both free and reference layer, appears to stay constant. P-to-AP switching could be observed for $t_{\mathrm{Cu}}$ up to 20 nm.

\begin{table*}
    \centering
    \caption{Thermophysical parameters used in the calculation taken from Refs. \cite{remy2023accelerating,karki2000high,lee1995thermal}}
    \begin{tabular}{lcccccc}
        \hline\hline
        & \textbf{Ta} & \textbf{Cu} & \textbf{MgO} &\textbf{[Co/Pt]} & \textbf{Pt} & \textbf{Glass} \\
        \hline
        $\gamma$ (J m$^{-3}$ K$^{-2}$) & 543  & 98 & -- & 720 & 749 & -- \\
        $C_{\mathrm{ph}}$ ($10^6$ J m$^{-3}$ K$^{-1}$) & 2.23 & 2.63 & 3.35 & 2.98 & 3.45 & 2 \\
        $\kappa_\mathrm{e}$ (W m$^{-1}$ K$^{-1}$) & 58 & 300 & -- & 20 & 45 & -- \\
        $\kappa_{p}$ (W m$^{-1}$ K$^{-1}$) & 5 & 5 & 4 & 1 & 5 & 2  \\
        $g_{\mathrm{ep}}$ (PW m$^{-3}$ K$^{-1}$) & 1000 & 75 & -- & 264 & 1100 & -- \\
        \hline
    \end{tabular}
    \label{tab1}
\end{table*}

\begin{figure*}
    \centering
    \includegraphics[width=0.8\linewidth, clip]
    {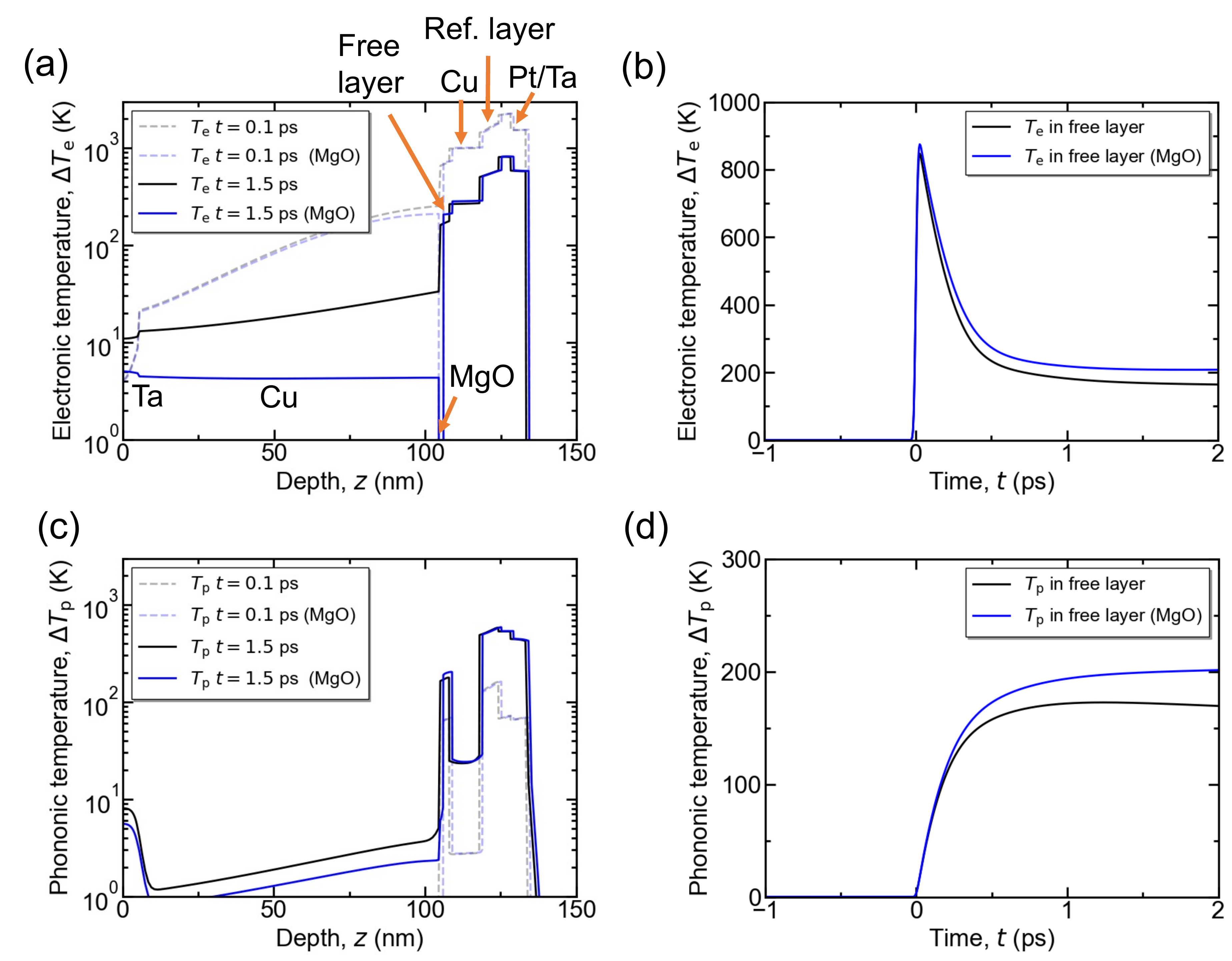}
    \caption{Thermal calculations based on a two-temperature model in Ta(5)/Pt(4)/[Co(0.82)/Pt(1)]$_{3}$/Co(0.82)/Cu(10)\allowbreak/[Co(0.49)/Pt(1)]$_{2}$/ MgO($t_{\mathrm{MgO}}$=0, 1)/Cu(100)/Ta(5). Spatial (a) electronic and (c) phononic temperature change in samples with $t_{\mathrm{MgO}}$=0 (black), 1(blue), respectively. Dashed and solid lines represent the change in the temperature at the time of 0.1 ps and 1.5 ps. Time evolution of (b)electronic and (d)phononic temperature change in the free layer in the sample with and without MgO layer.}
    \label{EM_fig3}
\end{figure*}

\textit{Appendix C: Thermal calculations.} In order to elucidate the effect of MgO insertion on heating, we performed the thermal calculations based on a two-temperature model using the following equations \cite{pudell2020heat}:

\begin{align}
C_{\mathrm{e}}(T_{\mathrm{e}})\frac{\partial T_{\mathrm{e}}}{\partial t} &= g_{\mathrm{ep}} (T_\mathrm{p} - T_\mathrm{e})+\nabla_z (\kappa_{\mathrm{e}} \nabla_z T_\mathrm{e}) + Q(z,t),\label{TC_electron}\\
C_{\mathrm{p}}\frac{\partial T_{\mathrm{p}}}{\partial t} &= g_{\mathrm{ep}} (T_\mathrm{e} - T_\mathrm{p}) +\nabla_z (\kappa_{\mathrm{p}} \nabla_z T_\mathrm{p}),
\label{TC_phonon}
\end{align}

where $T_{\mathrm{e}}$ and $T_{\mathrm{p}}$ are the electronic/phononic temperatures, respectively. $C_{\mathrm{e/p}}$ and $\kappa_{\mathrm{e/p}}$ represent electronic/phononic specific heats and thermal conductivities, respectively. $g_{\mathrm{ep}}$ is the coupling constant between electrons and phonons. $Q$ represents the heat source associated with the laser pulse excitation from the substrate side. Here, the electronic specific heat is given by $C_{\mathrm{e}}=\gamma T_{\mathrm{e}}$, where $\gamma$ represents the Sommerfeld constant. The parameters used for thermal calculations are listed in Tab. \ref{tab1}.

Fig. \ref{EM_fig3}(a) and (c) show the spatial profile of electronic and phononic temperatures at the delay time of 0.1 ps and 1.5 ps. Black and blue lines represent the calculation in the sample without and with the MgO layer, respectively. Since  MgO blocks electron transport, the upper Cu and Ta layers are more heated in the sample without the MgO layer. Furthermore, it was found that the electronic and phononic temperatures in the free layer are higher in the sample with MgO. The time evolutions of $T_\mathrm{e}$ and $T_\mathrm{p}$ in the free layer are shown in Fig. \ref{EM_fig3}(b) and (d), respectively. The temperature difference between two samples is gradually extended. Note that the electronic and phononic temperatures in other layers are almost the same, even in the reference layer. Therefore, the thermal calculations reveal that the MgO insertion affects the electronic and phononic temperatures in the spin valves, implying that control of heating in the free layer is one of the key factors in P-to-AP switching.

\end{document}